\documentclass[trackchanges,twocolumn]{aastex7}
\usepackage{comment}

\begin{document}

\title{Binary Constraints on the Origin of Nitrogen-rich Field Stars}

\author[orcid=0000-0000-0000-0001]{Liao Yang}
\affiliation{Department of Astronomy, School of Physics and Astronomy, Sun Yat-sen University, Zhuhai, Guangdong Province, China}
\affiliation{CSST Science Center for the Guangdong-Hong Kong-Macau Greater Bay Area, Zhuhai 519082, China}
\email{yangliao@mail2.sysu.edu.cn}  

\author[orcid=0000-0002-0066-0346]{Baitian Tang} 
\affiliation{Department of Astronomy, School of Physics and Astronomy, Sun Yat-sen University, Zhuhai, Guangdong Province, China}
\affiliation{CSST Science Center for the Guangdong-Hong Kong-Macau Greater Bay Area, Zhuhai 519082, China}
\email[show]{tangbt@mail.sysu.edu.cn}

\author[orcid=0000-0003-3526-5052]{Jos\'e G. Fern\'andez-Trincado} 
\affiliation{Centro de Investigaci\'on en Astronom\'ia, Universidad Bernardo O’Higgins, Avenida Viel 1497, Santiago, Chile}
\email{jose.fernandez@ucn.cl}

\author[orcid=0000-0002-3084-5157]{Chengyuan Li} 
\affiliation{Department of Astronomy, School of Physics and Astronomy, Sun Yat-sen University, Zhuhai, Guangdong Province, China}
\affiliation{CSST Science Center for the Guangdong-Hong Kong-Macau Greater Bay Area, Zhuhai 519082, China}
\email{lichengy5@mail.sysu.edu.cn}

\author[orcid=0000-0001-8713-0366]{Long Wang} 
\affiliation{Department of Astronomy, School of Physics and Astronomy, Sun Yat-sen University, Zhuhai, Guangdong Province, China}
\affiliation{CSST Science Center for the Guangdong-Hong Kong-Macau Greater Bay Area, Zhuhai 519082, China}
\email{wanglong8@mail.sysu.edu.cn}

\author[orcid=0000-0003-4265-7783]{Dengkai Jiang}
\affiliation{International Centre of Supernovae (ICESUN), Yunnan Key Laboratory of Supernova Research, Yunnan Observatories, Chinese Academy of Sciences (CAS), Kunming 650216, China}
\email{dengkai@ynao.ac.cn}

\author[orcid=0000-0002-0378-2023]{Bo Ma} 
\affiliation{Department of Astronomy, School of Physics and Astronomy, Sun Yat-sen University, Zhuhai, Guangdong Province, China}
\affiliation{CSST Science Center for the Guangdong-Hong Kong-Macau Greater Bay Area, Zhuhai 519082, China}
\email{mabo8@mail.sysu.edu.cn}

%\author{River Europe}
%\affiliation{University of Heidelberg}
%\email{fakeemail4@google.com}

%\author[0000-0000-0000-0003,sname=Asia,gname=Mountain]{Asia Mountain}
%\altaffiliation{Astrosat Post-Doctoral Fellow}
%\affiliation{Tata Institute of Fundamental Research, Department of Astronomy}
%\email{fakeemail5@google.com}

%\author[0000-0000-0000-0004]{Coral Australia}
%\affiliation{James Cook University, Department of Physics}
%\email{fakeemail6@google.com}

%\author[gname=IceSheet]{Penguin Antarctica}
%\affiliation{Amundsen–Scott South Pole Station}
%\email{fakeemail7@google.com}

\begin{abstract}
Recent JWST observations have revealed galaxies with unusually high N/O ratios, suggesting that nitrogen enrichment may be common in intense star-forming environments in the early Universe. In the Milky Way, nitrogen-rich(N-rich) stars in the Galactic field have long served as probes of early Galaxy formation and globular cluster enrichment. However, the identification of binaries among these stars raises the possibility that binary mass transfer could contribute to their origin.
In this work, we utilize multi-epoch radial velocities and element abundances from APOGEE DR17 to constrain their formation sites. Among 266 N-rich field stars, 33 exhibit radial velocity variations of $\Delta {\rm RV} > 1\,{\rm km/s}$, including 10 robust spectroscopic binaries identified using the $F_2$ statistic within a well-sampled subset of 46 stars. The resulting close-binary fraction ($21.7\pm6.1\%$) is statistically indistinguishable from that of chemically normal field stars ($18.1\pm0.6\%$), showing no evidence of the excess expected from AGB binary pollution.  This is further supported by the absence of correlation between [N/Fe] and [Ce/Fe] and the lack of [C/Fe] enhancement. Crucially, we detect an anti-correlation between binary fraction and [Al/Fe], with strongly Al-enhanced stars ($[\mathrm{Al/Fe}] \gtrsim 0.5$) exhibiting a reduced binary fraction ($< 10\%$). This trend serves as a dynamical fingerprint of high-density environments, consistent with the efficient disruption of binaries via three-body interactions in GC cores. Our results do not support binary mass transfer as the dominant formation channel for N-rich field stars;  they are predominantly GC escapees that retain the dynamical memory of their dense birth sites.
\end{abstract}

%Furthermore, theoretical expectations predict an occurrence rate nearly two orders of magnitude below the observed frequency.
%%\keywords{nitrogen-rich}


\section{Introduction}\textbf{}

Recent \textit{JWST} observations have revealed a population of high-redshift galaxies exhibiting unusually strong nitrogen emission, including the remarkable case of GN-z11 at $z \sim 11$ \citep{Cameron2023, Marques-Chaves2024}. The extreme N/O abundance ratios inferred in these systems are difficult to explain within conventional chemical evolution models and may instead point to enrichment channels associated with dense stellar environments, such as runaway stellar collisions, supermassive stars, or the early formation of globular cluster (GC)-like systems. These discoveries suggest that nitrogen enhancement may serve as a tracer of intense clustered star formation in the early Universe.

Nearby nitrogen-rich field stars in the Milky Way may therefore provide a local fossil record of similar enrichment processes. Large spectroscopic surveys have uncovered field stars exhibiting enhanced N abundances together with GC-like light-element patterns \citep{Martell2011, Martell2016, Schiavon2017, Fernandez-Trincado2022}. Because these abundance signatures resemble those of second-generation (SG) stars in Galactic GCs, many studies have interpreted N-rich field stars as remnants of disrupted clusters dispersed into the Galactic halo and bulge \citep{Tang2019, Horta2021, Belokurov2023, Xu2024}. Several works have also reported that N-rich field stars are more common in the inner Milky Way, suggesting a possible connection to dense birth environments and GC-related formation channels\citep{Horta2021,Belokurov2023,Kane2026}. If true, these stars would provide an important local probe of early clustered star formation and GC assembly across cosmic time.

Besides the widely accepted GC escapee scenario, binary mass transfer from AGB companions has been proposed as an alternative mechanism for producing nitrogen enhancement in field stars \citep[e.g.,][]{Martell2016,Schiavon2017}. In particular, \citet{FernandezTrincado2019b} discovered an individual N-rich star exhibiting large radial-velocity variations and signatures of a binary companion, suggesting that it may have formed through binary mass transfer from an AGB companion. %These results raise the question of whether binary mass transfer represents an important contribution to the N-rich field-star population or is limited to a small fraction of individual systems.
%Similarly, the extremely nitrogen-enhanced metal-poor (NEMP) stars discussed by \citet{Simpson2019} are generally interpreted as products of binary evolution involving AGB companions. However, these systems exhibit much more extreme nitrogen enrichment than typical N-rich field stars and are thought to represent a distinct population. Conversely, the presence of Al enhancement and GC-like light-element abundance patterns in many N-rich field stars has been widely interpreted as evidence for an origin in disrupted globular clusters . 
However, AGB-based self-enrichment has also been widely discussed as a possible origin for multiple stellar populations within GCs themselves \citep{Dantona2002, DErcole2008, Ventura2013, Bastian2018}. Consequently, purely chemical diagnostics alone may not uniquely distinguish between GC-origin and binary-pollution scenarios. 

Binary properties provide a fundamentally different and more direct test of the mass-transfer hypothesis. If a substantial fraction of N-rich stars originated from binary pollution, they should exhibit enhanced radial-velocity variability and potentially correlations between N enrichment and carbon or $s$-process abundances. The abundance of Al is also an important diagnostic, since Al-enhanced field stars have already been found and are often interpreted in the context of GC-like multiple populations\citep{FernandezTrincado2020,Gratton2012}.
The detection of N-rich field stars with significant radial-velocity variability underscores the pressing need to quantify binary mass transfer, motivating a dedicated multi-epoch radial velocity survey of a large sample.
In this work, we combine APOGEE DR17 multi-epoch radial velocities with chemical abundance information to systematically examine the binary nature of Galactic N-rich field stars for the first time. We show that only a small fraction of N-rich stars exhibit strong binary signatures, significantly below expectations from a dominant AGB mass-transfer origin. In addition, the absence of strong correlations between N enrichment and Ce enhancement disfavors widespread pollution from transition-mass AGB companions. These results provide new evidence that most N-rich field stars are more naturally linked to GC-like stellar populations rather than binary evolution.

\section{DATA AND SAMPLE SELECTION}

We use high-resolution near-infrared spectra and multi-epoch radial velocities from APOGEE DR17 \citep{Abdurrouf2022}. APOGEE provides $H$-band ($R\sim22,500$) spectroscopy together with repeat radial-velocity (RV) measurements, making it well suited for identifying spectroscopic binaries through RV variability \citep{Nidever2015}.

To investigate the origin of nitrogen-rich (N-rich) field stars, we construct three comparison samples:

\begin{itemize}
    \item \textbf{Group A (N-rich field stars):} our primary sample consists of N-rich stars from \citet{Fernandez-Trincado2022}. They exhibit significant enrichment in nitrogen abundance ratios ([N/Fe] $>$ +0.5), along with simultaneous depletions in their [C/Fe] abundance ratios ([C/Fe] $<$ +0.15). Stars associated with the Magellanic Clouds were excluded to focus on the Galactic population.
    
    \item \textbf{Group B (N-rich GC stars):} globular cluster (GC) members with [N/Fe] $\geq +0.5$ selected from the catalog of \citet{Schiavon2024}, which identifies GC members using spatial positions, radial velocities, and metallicities.
    
    \item \textbf{Group C (N-normal field stars):} APOGEE field stars with [N/Fe] $< +0.5$. Possible GC members were removed by excluding stars located within 1.5 tidal radii, $\pm0.3$ dex in metallicity, and $\pm15$ km s$^{-1}$ in RV relative to known GCs.
\end{itemize}

We applied uniform quality cuts to all samples. Stars with only one visit or signal-to-noise ratio $S/N < 5$ were excluded, and we required at least two RV measurements for binary analysis. To ensure homogeneous stellar-parameter coverage, Groups B and C were restricted to the parameter range of the N-rich sample:
\[
3800\,{\rm K} < T_{\rm eff} < 5350\,{\rm K}, \quad
0.08 < \log g < 3.25.
\]

Because binary fraction correlates with metallicity \citep{Moe2019}, we further matched the metallicity distribution of Group C to that of Group A through weighted resampling. For each metallicity bin of the N-rich sample, we randomly selected 40 times more control stars from Group C to minimize statistical noise while preserving the same [Fe/H] distribution.

The final samples contain 266 N-rich field stars (Group A), 2601 N-rich GC stars (Group B), and a resampled control sample of $\sim 10,\!280$ N-normal field stars (Group C). The visit-number and temporal-baseline distributions of the three samples are broadly similar (Appendix~\ref{appendix:visit}).

\section{Binary Identification}
\label{F2}
The multi-epoch APOGEE observations naturally provide time-domain radial-velocity (RV) information that can be used to identify spectroscopic binaries. However, most APOGEE targets have only a limited number of visits, making robust binary identification challenging.

To quantify RV variability while accounting for different numbers of observations, we adopt the $F_2$ statistic \citep{Jofre2016,Jorissen2020,Jofre2023}. We first compute the RV $\chi^2$ statistic relative to the mean velocity:

\[
\chi_j^2=\sum_i \frac{(RV_{i,j}-\langle RV_j\rangle)^2}{\sigma^2},
\]

with degrees of freedom $\nu_j = N_{\text{visit}} - 1$, where $\sigma$ denotes an adopted characteristic RV uncertainty for a given star. In this paper, $\sigma$ is treated as a constant for each source, reflecting the long-term instrumental stability of APOGEE rather than visit-level nominal uncertainties. Specifically, APOGEE RV measurements typically have nominal uncertainties of $\sim 0.1 - 0.2$ km/s \citep{Nidever2015}. The APOGEE-provided uncertainties decrease with increasing metallicity and SNR. To mitigate spurious detections caused by occasional low-S/N visits and intrinsic stellar RV jitter, we conservatively adopt an inflated uncertainty of $\sigma = 0.4$ km$^{-1}$ in our analysis.

The reduced $\chi^2$ is then transformed into the approximately Gaussian-distributed $F_2$ statistic \citep{WilsonHilferty1931,CastroTapia2024}:

\[
F_{2,j}=
\sqrt{\frac{9\nu_j}{2}}
\left[
\left(
\frac{\chi_j^2}{\nu_j}
\right)^{1/3}
+\frac{2}{9\nu_j}-1
\right].
\]

Under the null hypothesis of constant RV, $F_2$ approximately follows a standard normal distribution. Large values therefore indicate statistically significant RV variability. We adopt $F_2>3$ as the criterion for identifying candidate spectroscopic binaries, corresponding approximately to a $3\sigma$ detection threshold.

Since the reliability of the $F_2$ statistic decreases for sparsely sampled stars, we further limit the three samples with $N_{\rm visit}\geqslant6$. The three sub-samples sizes are 46, 719, and 4198 for Group A$_{N\geqslant6}$, B$_{N\geqslant6}$, and C$_{N\geqslant6}$, respectively.\textbf{ }We confirmed that the metallicity distribution consistency between the Group A$_{N\geqslant6}$ and C$_{N\geqslant6}$ sub-samples are preserved.\textbf{
}

\section{Results}
\label{sec:results}
\subsection{Binary Fraction}
\label{sec:binary_fraction}

Applying  $F_2$ criterion, we derive the observed close-binary fractions for the three samples, which are summarized in Table~\ref{tab:binary_fractions}. The N-rich field stars exhibit a binary fraction that is statistically indistinguishable from that of N-normal field stars within the quoted uncertainties. In contrast, the globular cluster (GC) sample shows a systematically lower close-binary fraction than both field samples.
To account for observational selection effects, we correct the observed binary fractions using the detection efficiency $\epsilon$ and false-positive rate $f_{\rm FP}$ derived from forward Monte Carlo simulations (Appendix~\ref{appendix:bf correction}). We find that all three samples exhibit nearly identical detection efficiencies ($\epsilon \simeq 0.9$) and extremely low false-positive rates ($f_{\rm FP} \simeq 10^{-3}$). This demonstrates that differences in observing cadence, visit number, and time baseline among the samples do not significantly bias the identification of close binaries, and therefore do not drive the observed trends in binary fraction.

\begin{table*}
\centering
\caption{Observed and corrected close-binary fractions}
\label{tab:binary_fractions}
\begin{tabular}{lcccccc}
\hline
Sample & $N_{\rm bi}$ & $N_{\rm tot}$ & $f_{\rm obs}$ (\%) & $\epsilon$ & $f_{\rm FP}$ & $f_{\rm true}$ (\%) \\
\hline
N-rich Field      & 10  & 46   & $21.74 \pm 6.08$ & $0.928$ & $0.001$ & 23.36 \\
Normal Field     & 761 & 4198 & $18.13 \pm 0.59$ & $0.909$ & $0.001$ & 19.87\\
N-rich GC         & 107 & 719  & $14.88 \pm 1.33$ & $0.928$ & $0.001$ & 15.94 \\
\hline
\end{tabular}
\end{table*}

We list all N-rich field stars identified as binary candidates in the first ten rows of Table~\ref{tab:nrich_binaries}. These systems represent the subset of N-rich field stars with clear radial-velocity variability and provide direct observational evidence that binaries do exist among them. We test their formation scenarios below by comparing with other two subsamples:

\begin{itemize}
    \item The lack of a statistically significant enhancement in the close-binary fraction of N-rich field stars relative to N-normal field stars is a key observational result. If the majority of N-rich stars originated from binary interaction channels involving intermediate-mass asymptotic giant branch (AGB) companions, one would generically expect a higher incidence of close binaries among the N-rich population (e.g., \citealt{Raghavan2010, Moe2017}). Our results do not support this expectation.
    \item The close-binary fraction measured in the GC sample is lower than that of both field samples. This behavior is consistent with long-standing theoretical expectations and observational studies showing that binaries in dense stellar environments are efficiently disrupted through dynamical interactions (e.g., \citealt{Hut1992, Ivanova2005, Milone2012}). Hard--soft interactions and close encounters preferentially destroy wide and intermediate-separation binaries over a Hubble time. If N-rich stars originate in globular clusters and escape into the field, their observed close-binary fraction may reflect the binary population at the time of ejection for at least a subset of stars, rather than the present-day GC binary fraction. Numerical simulations by \citet{Weatherford2023} show that some stars can escape from globular clusters at early evolutionary stages. In this scenario, escaped N-rich stars may retain a close-binary fraction comparable to that of typical field stars, consistent with our observational results.
\end{itemize}

\begin{table*}
\centering
\caption{N-rich field stars with N$_{visit}\geqslant6$}
\label{tab:nrich_binaries}
\begin{tabular}{ccccccc}
\hline
APOGEE\_ID & RA (deg) & Dec (deg) & $\Delta RV$ (km s$^{-1}$) & $N_{\rm visit}$ & $F_2$ & Reference DOI \\
\hline
2M18211586$-$3856467 & 275.316101 & $-$38.946308 & 10.24 & 6  & 21.98 &10.1093/mnras/stab525\\
2M18364041$-$3402389 & 279.168375 & $-$34.044147 & 12.21 & 7  & 18.91 & 10.1093/mnras/stab525 \\
2M12451043$+$1217401 & 191.293486 & 12.294486  & 20.62 & 25 & 74.67 & 10.1051/0004-6361/201935369 \\
2M02000451$-$0229333 & 30.018804  & $-$2.492584 & 16.87 & 10 & 42.60 & 10.1093/mnras/stz1848 \\
2M15082716$+$6710075 & 227.113188 & 67.168762  & 3.55  & 45 & 9.76  & 10.1093/mnras/stz1848 \\
2M19193412$-$2931210 & 289.892195 & $-$29.522511& 6.02  & 6  & 11.24 & 10.1051/0004-6361/202039434 \\
2M18565969$-$3106454 & 284.248713 & $-$31.112616& 2.43  & 12 & 4.27  & 10.1051/0004-6361/202140306 \\
2M06243113$+$4213097 & 96.129723  & 42.219379  & 5.41  & 8  & 8.89  & 10.1051/0004-6361/202346325 \\
2M01502296$+$1341056 & 27.595676  & 13.684916  & 16.01 & 7  & 32.26 & 10.1051/0004-6361/202346325 \\
2M01575297$-$0316508 & 29.470726  & $-$3.280803 & 18.11 & 7  & 38.96 & 10.1051/0004-6361/202346325 \\
...&...&...&...&...&...&...\\
\hline
\end{tabular}
\medskip
\\ \raggedright \hspace{0.5 cm}\small{Full machine-readable table will be provided online.}
\end{table*}

\begin{figure*}
    \centering
    \includegraphics[width=0.75\textwidth]{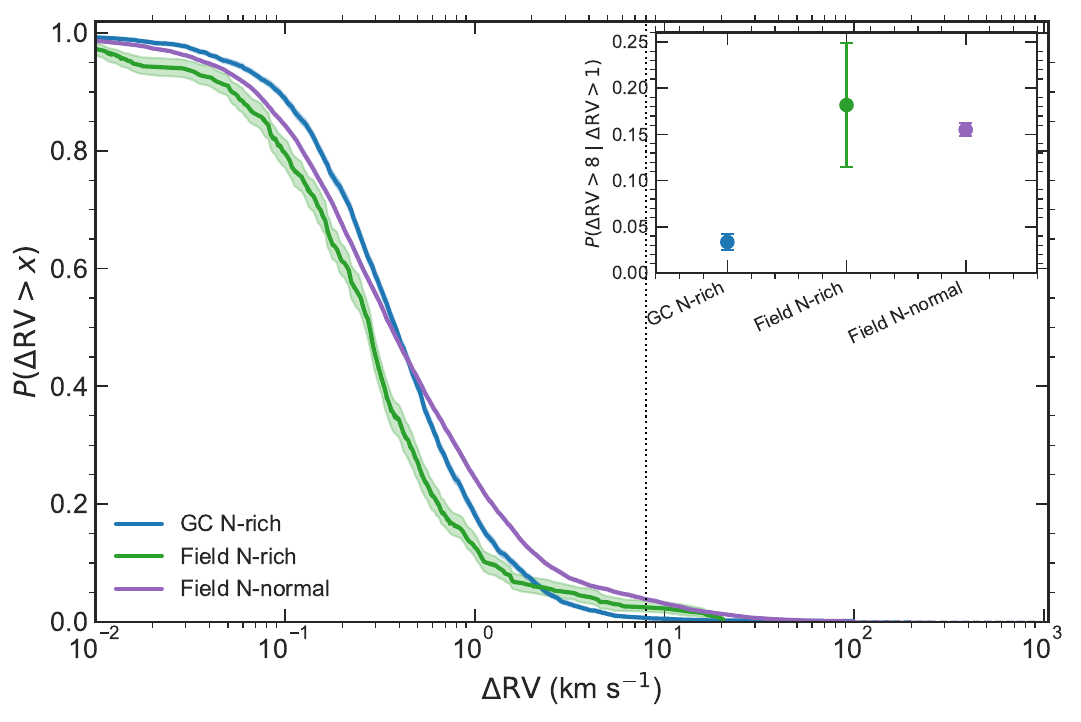}
    \caption{
Cumulative distributions of radial-velocity variability for different stellar samples.
Solid curves show the median cumulative distributions from bootstrap resampling for N-rich field stars (green), GC stars (blue), and normal field stars (purple). The shaded region indicates the 16th–84th percentile range for the N-rich field sample. The vertical dotted line marks $\Delta RV=8$ km s$^{-1}$.
Inset: the fraction of stars with $\Delta RV>8$ km s$^{-1}$(extreme systems) among all stars with $\Delta RV>1$ km s$^{-1}$(binary candidates),
$P(\Delta RV > 8~{\rm km~s^{-1}} \mid \Delta RV > 1~{\rm km~s^{-1}})$}.

    \label{fig:rv_ccdf}
\end{figure*}

Beyond the overall binary fraction, the distribution of radial-velocity variability provides additional constraints on the nature of binary systems. Large $\Delta RV$ values are expected for compact binaries with short orbital periods, which are natural outcomes of unstable Roche-lobe overflow or common-envelope evolution involving intermediate-mass AGB companions.

The cumulative distributions of $\Delta RV$ (Figure~\ref{fig:rv_ccdf}) suggested that N-rich field stars do not exhibit a pronounced high-$\Delta RV$ tail compared to either GC N-rich stars or N-normal field stars. The inset in Figure~\ref{fig:rv_ccdf} summarizes the fraction of stars with $\Delta RV>8$ km s$^{-1}$(extreme systems) among all stars with $\Delta RV>1$ km s$^{-1}$(binary candidates).\textbf{ }This threshold ($8$ km/s) corresponds to orbital separations of a few tenths of an AU for typical giant-star binaries, where unstable Roche-lobe overflow or common-envelope evolution is expected \citep{Hansen2016}. As shown in the inset figure, the corresponding fractions for the N-rich stars and the normal field stars are comparable.
The absence of an excess population of such extreme systems among N-rich field stars further disfavors a dominant binary mass-transfer origin for this population.

\subsection{Chemical Properties of Binary N-rich Stars}
\label{sec:abundance}

Previous studies have already established that N-rich field stars exhibit chemical abundance patterns similar to multiple-population stars in globular clusters (GCs), including enhanced N and Al together with depleted C \citep{Yu2021, Fernandez-Trincado2022,Qiao2026}. Such abundance signatures are widely interpreted as evidence for a GC origin. Interestingly, N-rich field stars have, on average, lower nitrogen abundances than GC stars. This trend likely reflects the higher average metallicity of field stars, given that N production is modulated by extra-mixing processes whose efficiency scales with metallicity \citep{Lagarde2012}. In contrast, the metal-poor N-rich field stars studied by \citet{Qiao2026} display N-enhancement levels that are comparable to those of canonical GC stars.

\begin{figure}[h!]
    \centering
    \includegraphics[width=\linewidth]{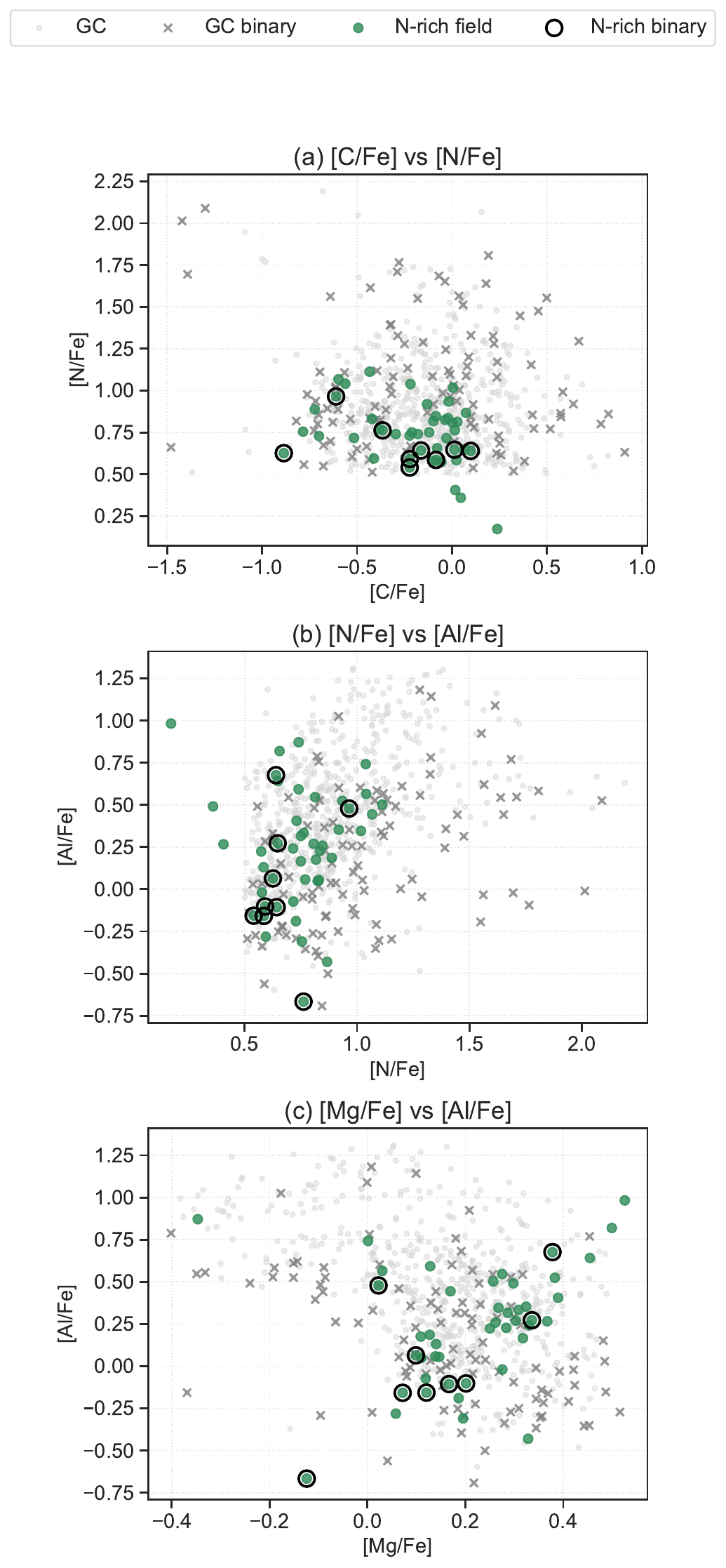}
    \caption{
    Chemical abundance distributions of N-rich field stars and GC stars. 
    Panels show (a) [C/Fe] versus [N/Fe], (b) [N/Fe] versus [Al/Fe], and (c) [Mg/Fe] versus [Al/Fe]. 
    Binary candidates are highlighted with black circles.
    }
    \label{fig:abundance}
\end{figure}

Here we focus specifically on whether the binary N-rich stars show chemical properties distinct from the non-binary population (Figure~\ref{fig:abundance}). %compares the abundance distributions of binary and non-binary N-rich stars in [C/Fe]--[N/Fe], [N/Fe]--[Al/Fe], and [Mg/Fe]--[Al/Fe] space.
We find no clear chemical distinction between the binary and non-binary subsamples among the abundances of C, N, Mg, and Al. The binary systems span the same abundance space as the rest of the N-rich population, indicating that the observed chemical anomalies are not uniquely associated with binary evolution or mass transfer. This result is consistent with the absence of an enhanced close-binary fraction among N-rich field stars.

Taken together, the chemical similarity between binary and non-binary N-rich stars further supports the interpretation that the majority of N-rich field stars originate from GC-like environments rather than from binary mass-transfer channels.

\section{Discussion} \label{sec:discussion}

\subsection{Statistical Occurrence: Challenging the Binary Pollution Scenario} \label{subsec:statistics}

A key constraint on the origin of N-rich stars is provided by their observed occurrence rate in the Galactic field. Previous large-scale spectroscopic surveys have consistently shown that nitrogen-enhanced field stars constitute a non-negligible fraction of the metal-poor population, with an incidence of approximately 1--17\% among  galactic field giants \citep{Martell2011,  Schiavon2017, Tang2020, Horta2021}. Importantly, this incidence strongly varies with Galactic radius: it is $>10\%$ in the inner Galaxy ($R\sim2\text{--}3$ kpc), while dropping to only a few percent at $R\sim5\text{--}10$ kpc\citep{Horta2021,Belokurov2023,Kane2026}. If these stars were produced via mass transfer from a more massive AGB companion in a binary system---analogous to the formation of barium or CH stars---the theoretical occurrence rate should be commensurate with the product of the Initial Mass Function (IMF), binary fraction, and the survival probability of such systems.

To test this hypothesis, we perform a first-order estimate of the N-rich star formation probability in the field, $P_{\rm N-rich, field}$. The enrichment of nitrogen on the surface of a lower-mass recipient requires a donor star in the intermediate-mass range ($4\,M_\odot \lesssim M \lesssim 8\,M_\odot$), as these stars undergo hot-bottom burning (HBB) during the AGB phase \citep{Karakas2014}. According to a standard \citet{Kroupa2001} IMF, the fraction of stars born within this mass range is approximately $f_{\rm IMF} \approx 1\%$. While the binary fraction for intermediate-mass stars is relatively high, $f_{\rm bin} \approx 50\%$--$70\%$ \citep{Moe2017}, the critical bottleneck is the survival probability through the mass-transfer phase ($f_{\rm surv}$).

Mass transfer from an AGB star to a lower-mass companion is frequently unstable due to the high mass ratio and the deep convective envelope of the donor. This instability often triggers common-envelope (CE) evolution, which typically results in a stellar merger or the formation of a very short-period binary system \citep{Ivanova2013}. Stable enrichment of the companion requires either stable Roche-lobe overflow or efficient wind accretion within a restricted range of orbital configurations. Binary evolution studies indicate that only a small fraction of interacting systems can avoid unstable mass transfer and remain long-lived binaries \citep{Pavlovskii2015,Abate2015}. Adopting a conservative order-of-magnitude estimate, we therefore take the survival probability for such systems to be $f_{\rm surv} \lesssim 10\%$. Combining these factors, we derive a theoretical occurrence rate:
\begin{equation}
    P_{\rm theory} \approx f_{\rm IMF} \times f_{\rm bin} \times f_{\rm surv} \approx 0.01 \times 0.7 \times 0.1 \approx 0.0007.
\end{equation}
This predicted frequency ($\sim 0.07\%$) is nearly two orders of magnitude lower than the observed 1\%--17\% frequency of N-rich stars in the galactic field. This large discrepancy strongly disfavors binary mass transfer from intermediate-mass AGB companions as the dominant formation channel of N-rich field stars.

We emphasize that this estimate is intended only as an order-of-magnitude calculation with several possible improvements. First, it does not explicitly account for the mass-ratio distribution of interacting binaries, which would likely reduce the effective formation rate for present-day low-mass ($\sim0.8\,M_\odot$) N-rich stars. Second, binary population synthesis models are known to underproduce observed Ba/CH/CEMP-$s$ systems and to poorly reproduce their observed period--eccentricity distributions (e.g., Abate et al. 2013, 2018; Escorza et al. 2019; Temmink et al. 2023), suggesting that simplified prescriptions for stable mass transfer may underestimate the number of surviving interacting binaries. In particular, wind-assisted Roche-lobe overflow (WRLOF) has been identified as an important mass-transfer channel that can substantially increase the efficiency of wind accretion in wide binaries. Although these effects may change the quantitative estimate, they are unlikely to eliminate the discrepancy of approximately one to two orders of magnitude between the expected occurrence rate of the binary mass-transfer channel and the observed incidence of N-rich field stars. Therefore, our qualitative conclusion remains unchanged.

\subsection{Dynamical Imprints on the Binary Fraction} \label{subsec:dynamics}

The binary fraction ($f_{\rm bin}$) of a stellar population serves as a fossil record of its dynamical history. While chemical abundances reveal the nucleosynthetic sites of star formation, the survival rate of binary systems encodes information about the stellar density and gravitational interactions experienced by the population. We will discuss the binary fraction as a function of metallicity ([Fe/H]) and aluminum abundance ([Al/Fe]) in this section.
Since the number of $F_2$-identified N-rich field binaries with $N_{\rm visit} \geqslant6$ is only 10, which is insufficient for such discussion, we use a more generous criterion for binary identification: $\Delta {\rm RV} > 1\,{\rm km/s}$. Considering APOGEE's nominal uncertainty of $\sim 0.1-0.2$ km/s and an inflated uncertainty of $\sigma=0.4$ km/s, the criterion of $\Delta {\rm RV} > 1$ km/s (2.5$\sigma$) is suitable to identify binaries. In the end, 33 stars are identified as binaries with $\Delta {\rm RV} > 1\,{\rm km/s}$ out of the Group A sample (266 stars).
%In Figure \ref{fig:BF_trends}, we investigate how the binary fraction (proxied by the fraction of stars with $\Delta {\rm RV} > 1\,{\rm km\,s^{-1}}$) varies with both metallicity ([Fe/H]) and aluminum abundance ([Al/Fe]).

\subsubsection{Absence of Metallicity Dependence}

\begin{figure*}
    \centering
    \includegraphics[width=0.7\textwidth]{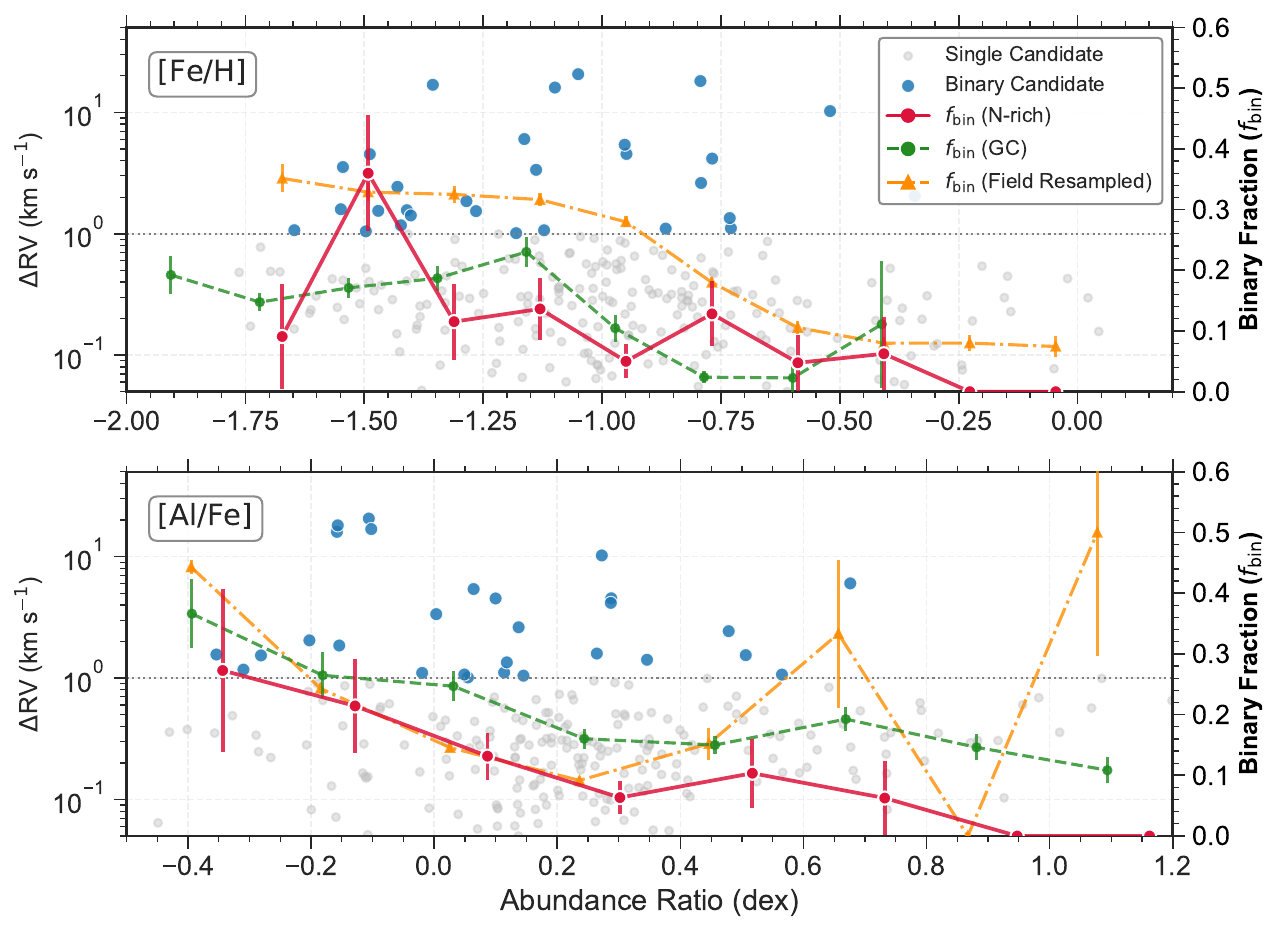} 
    \caption{The variation of radial velocity scatter ($\Delta$RV) of N-rich star and binary fraction of three sample as a function of [Fe/H] (top panel) and [Al/Fe] (bottom panel). Grey dots represent single candidates, while blue dots indicate binary candidates ($\Delta {\rm RV} > 1\,{\rm km\,s^{-1}}$). Curves with error bars represent the binned binary fractions for the GroupA (red), GroupB(green), and GroupC(orange).}
    
    \label{fig:BF_trends}
\end{figure*}

Figure~\ref{fig:BF_trends} compares the binary fraction as a function of metallicity for the N-rich field, normal field, and GC samples. The normal field stars exhibit a clear decrease in binary fraction toward higher metallicity, consistent with previous studies of Galactic field populations (e.g., \citealt{Yuan2015,Moe2019}). In contrast, both the N-rich field stars and the GC sample show no statistically significant metallicity dependence.

The contrasting behaviors among the three samples provide a direct internal comparison using the same binary-identification method and observational dataset, thereby minimizing systematic differences between independent studies. The agreement between the normal field sample and previous literature further demonstrates that our method successfully recovers the well-established metallicity dependence of the Galactic field population.

The absence of a comparable metallicity dependence among the N-rich field stars therefore appears to be intrinsic rather than an artifact of the analysis. Instead, their behavior closely resembles that of the GC sample. These suggests N-rich field stars have undergone a different evolutionary path compared to the normal field population. The primordial metallicity-binary relation, if present initially, may have been altered by the dense environments typical of globular clusters (GCs). In such environments, internal dynamical processing tends to dominate over initial formation conditions.The binary fraction trends we observe are consistent with the hypothesis that these N-rich stars are escaped members of massive clusters (either dissolved or surviving), where the intense dynamical interactions have effectively randomized or saturated the binary survival rates, erasing the memory of the initial metallicity-dependent fragmentation.

% --- Figure Caption ---

\subsubsection{Anti-correlation with [Al/Fe]}

The lower panel of Figure~\ref{fig:BF_trends} compares the dependence of binary fraction ($f_{\mathrm{bin}}$) on $\mathrm{[Al/Fe]}$ across the three samples. For the normal field population, $f_{\mathrm{bin}}$ displays a two-fold behavior: (1) a steep decline at low $\mathrm{[Al/Fe]}$ ($\mathrm{[Al/Fe]} \lesssim 0.3$), which reflects the transition in average metallicity between accreted stars (e.g., Gaia-Sausage-Enceladus; characterized by lower $\mathrm{[Al/Fe]}$, lower $\mathrm{[Fe/H]}$, and intrinsically higher $f_{\mathrm{bin}}$) and in-situ field populations \citep{Lin2025}; (2) no clear systematic trend with larger statistical uncertainties at higher abundances ($\mathrm{[Al/Fe]} \gtrsim 0.3$). Most MW field stars show [Al/Fe$]<0.4$ \citep{Tang2023}, and stars with higher [Al/Fe] should be considered as peculiar sources with GC origins or AGB pollutions. The observed behavior at high [Al/Fe] reflects the small sample size ($\sim$20 stars separated into three bins) and noneligible contribution from AGB companions. The latter is supported by their large [Ce/Fe] values --- more than half of the field stars with [Al/Fe$]>0.7$ show [Ce/Fe$]>0.5$. 

In contrast, both the N-rich field stars and the GC control sample maintain a consistently declining binary fraction trend across the entire $\mathrm{[Al/Fe]}$ range.
This is naturally expected if the N-rich stars originated in GC-like environments. In multiple-population formation scenarios, Al-rich stars preferentially form in the dense central regions of massive clusters \citep{DErcole2008,Decressin2007}, where frequent dynamical interactions efficiently disrupt primordial binaries \citep{Heggie1975,Vesperini2011,Hong2015}. Consequently, stars with stronger chemical enrichment are expected to retain lower surviving binary fractions.

The similarity between the N-rich and GC samples, together with their clear difference from the normal field population analyzed using the same methodology, provides further evidence that the binary properties of N-rich field stars are more closely linked to GC environments than to the general Galactic field.This interpretation is consistent with observations of present-day Milky Way GCs, where chemically enriched stars exhibit systematically lower binary fraction \citep{Lucatello2015,Dalessandro2022}.

\subsection{Chemical Signatures: Disentangling Nucleosynthesis Channels} \label{subsec:chemistry}

While the occurrence rates discussed in Section~\ref{subsec:statistics} provide statistical evidence against the dominant role of intermediate-mass AGB binaries, chemical abundance ratios offer an independent probe of the nucleosynthetic origin of the N-rich stars. 

From stellar evolution theory, strong nitrogen enhancement is primarily produced in intermediate-mass AGB stars ($\sim4$--$8\,M_\odot$) experiencing hot-bottom burning (HBB), whereas low-mass AGB stars mainly enrich their companions in carbon and $s$-process elements during third dredge-up episodes \citep{Busso1999, Karakas2014}. Although low-mass AGB stars are not expected to generate substantial nitrogen enhancement, stars near the transition mass range ($\sim3$--$4\,M_\odot$) may produce moderate N enrichment while still exhibiting accompanying C and $s$-process signatures. 

Therefore, if the N-rich field stars originate from binary mass transfer involving such AGB donors, one would expect correlated enhancements between  N and $s$-process elements (e.g., Ce). In contrast, second-generation globular cluster stars typically show enhanced N without corresponding $s$-process enrichment \citep{Gratton2012,Bastian2018}. The relationship between [N/Fe] and [Ce/Fe] thus provides a useful diagnostic to distinguish between binary mass transfer and globular cluster origin.

\begin{figure}
    \centering
    \includegraphics[width=0.47\textwidth]{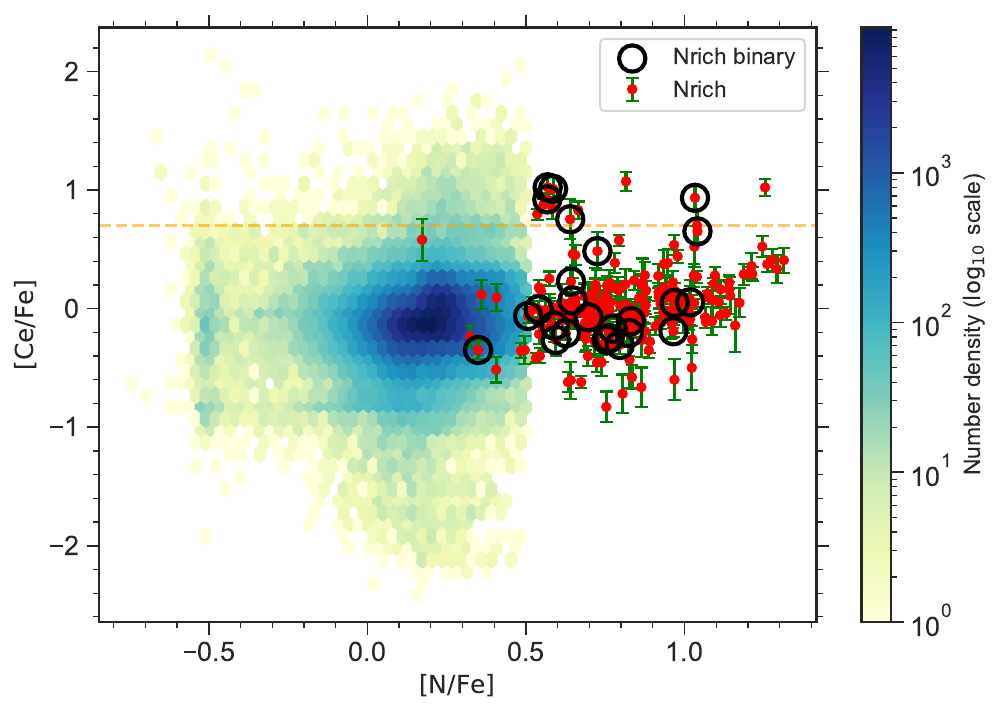}
    \caption{
Abundance correlations between [Ce/Fe] and [N/Fe] for the N-rich stars.
Red points denote N-rich stars, with open circles marking binary candidates identified from radial-velocity variability.
The background density map shows the parent APOGEE sample, and the dashed line indicates the adopted Ce-enhancement threshold. 
}
    \label{fig:N-Ce}
\end{figure}

To examine whether such correlation exists, we selected N-rich field stars with available [N/Fe] and [Ce/Fe] measurements from ASPCAP results. A sample of 237 stars were selected, with 29 identified binaries satisfying $\Delta {\rm RV} > 1\,{\rm km\,s^{-1}}$ 
As shown in (Figure \ref{fig:N-Ce}), the N-rich field stars do not follow a diagonal trajectory toward the upper right, as would be predicted by coupled enrichment models. While they are uniformly enhanced in nitrogen (${\rm [N/Fe]} \gtrsim 0.5$), their cerium abundances span a wide range ($-0.5 \lesssim {\rm [Ce/Fe]} \lesssim 1.0$) with no statistically significant correlation with nitrogen. A small number of binary N-rich stars do exhibit moderate Ce enhancement, suggesting that binary interaction may contribute to the formation of a minority of the population, particularly through transition-mass AGB donors. However, such objects represent only a small fraction of the N-rich sample and therefore cannot account for the population as a whole.
Furthermore, the N-rich field stars are selected without carbon enrichment (see Figure \ref{fig:abundance}a), at odds with that of typical CH stars. The absence of [C/Fe] enhancement, combined with the decoupling of [N/Fe] and [Ce/Fe] enrichment, suggests that the N-rich material did not originate from transition-mass AGB stars in binary systems. Instead, the abundance patterns---high N, low C, and variable s-process---are the hallmark of second-generation populations formed in GCs, where C is converted to N via the CNO cycle in the polluter's interior \citep{Gratton2004}.

\subsection{Caveats}

We emphasize that the derived binary fractions correspond specifically to short- and intermediate-period spectroscopic binaries ($P \lesssim 3-5$ years), which are accessible to APOGEE given its radial-velocity precision and temporal baselines. Our measurements do not probe the intrinsic binary fraction over all orbital separations, but rather the subset of close binaries that are most relevant for binary interaction channels such as mass transfer or common-envelope evolution. 

In addition, our study cannot detect binary systems that undergo mergers, though they may show chemical patterns similar to those of SG stars \citep{Kravtsov2025}. These merger remnants may no longer exhibit detectable radial-velocity variations. This may explain the apparently young ages inferred for some N-rich field stars with high metallicities \citep{Leitinger2026}.

\section{Summary} \label{sec:summary}

Since their discovery, nitrogen-rich (N-rich) field stars have been proposed to be the best candidates of GC escapees with similar chemical enrichment. However, the emerging binaries found in these types of stars have questioned their true formation sites. In this work, we have investigated the origin of N-rich field stars in the Milky Way by integrating their binary properties with chemical abundance patterns from APOGEE DR17. This work is part of our ongoing research project, ``Scrutinizing GAlaxy-STaR cluster coevolutiON with chemOdynaMIcs (GASTRONOMI)'', which leverages multi-wavelength photometric and spectroscopic data to unravel the coevolutionary relationships between the MW, its satellite dwarf galaxies, and their star clusters.

%By utilizing multi-epoch radial velocities and high-resolution spectroscopy from APOGEE DR17, we derive the close-binary fraction of N-rich field stars to provide a comprehensive constraint on the competing formation scenarios. Among our 266 N-rich field stars, 33 exhibit radial velocity variations of $\Delta RV > 1~{\rm km s^{-1}}$, including 10 robust close binaries identified via the $F_2$ statistic within the subset of 46 stars with $N_{\rm visit} \ge 6$.

Our main conclusions are as follows:

\begin{enumerate}
    \item \textbf{Constraints on the Binary Pollution Scenario:} We find that the close-binary fraction of N-rich field stars ($21.7 \pm 6.1\%$) is statistically consistent with that of the normal field population ($18.1 \pm 0.6\%$). The absence of a binary excess, combined with the low theoretical occurrence rate of stable mass-transfer systems,  disfavors intermediate-mass AGB companions as the primary formation channel for these stars. Furthermore, the observed decoupling between [N/Fe] and [Ce/Fe] abundances, coupled with the lack of [C/Fe] enhancement,  disfavors residual contributions from transition-mass AGB binaries, which would otherwise produce correlated C and $s$-process signatures.
    
    \item \textbf{Dynamical Evidence for GC Origin:} We observe a clear anti-correlation between the binary fraction and [Al/Fe] abundance. The depletion of binaries in the most Al-rich stars ($< 10\%$) serves as a dynamical fingerprint of high-density environments. This trend supports a scenario where N-rich stars are escapees from globular clusters, with the most Al-rich part having experienced the most intense dynamical processing and binary disruption within the dense cluster cores.
\end{enumerate}

Overall, our results demonstrate that the combination of binary properties and chemical tagging provides a powerful diagnostic for the origin of chemically peculiar stars. This paper argues against binary mass transfer as the dominant formation channel for N-rich field stars, while supporting disrupted or evaporated globular clusters as the most probable origin.

\begin{acknowledgements}
	$\hspace*{0.4cm}$L. Yang and B. Tang gratefully acknowledge support from the National Natural Science Foundation of China through grants NOs. 12473035 and 12233013, China Manned Space Project under grant NO. CMS-CSST-2025-A13 and CMS-CSST-2021-A08, the Fundamental Research Funds for the Central Universities, Sun Yat-sen University (24qnpy121). J.G.F-T gratefully acknowledges the grants support provided by ANID Fondecyt Regular No. 1260371, ANID Fondecyt Postdoc No. 3230001 (Sponsoring researcher), the Joint Committee ESO-Government of Chile under the agreement 2023 ORP 062/2023, and the support of the Doctoral Program in Artificial Intelligence, DISC-UCN. 
    L.W. thanks the support from the National Natural Science Foundation of China through grant 12233013 and 12573041, the GuangDong Basic and Applied Basic Research Foundation (2026A1515012460), the High-level Youth Talent Project (Provincial Financial Allocation) through the grant 2023HYSPT0706, the Fundamental Research Funds for the Central Universities, Sun Yat-sen University (2025QNPY04), and the support from the China Manned Space Project with NO.CMS-CSST-2021-A08.
    D.J. acknowledges support from the National Natural Science Foundation of China (Nos. 12473033, 12288102, 12333008, 12090040/3), the National Key R\&D Program of China No. 2021YFA1600403, International Centre of Supernovae, Yunnan Key Laboratory (No. 202302AN360001), Yunnan Revitalization Talent Support Program -- Science \& Technology Champion Project (NO. 202305AB350003), Yunnan Fundamental Research Project (No. 202201BC070003, 202401BC070007), the Yunnan Ten Thousand Talents Plan Young \& Elite Talents Project, ``Yunnan Revitalization Talent Support Program" for Young Talent Project, and the China Manned Space Project with No. CMS-CSST-2021-A08. 
    Large language models, including ChatGPT, DeepSeek, and Gemini, were used for language editing and polishing.
\end{acknowledgements}

\appendix

\setcounter{figure}{0} % 将图片计数器重置为0
\renewcommand{\thefigure}{A\arabic{figure}} % 定义图片编号显示格式为 A1, A2...

\section{APOGEE Visit Statistics}
\label{appendix:visit}

\begin{figure}[h!]
    \centering
    \includegraphics[width=0.7\linewidth]{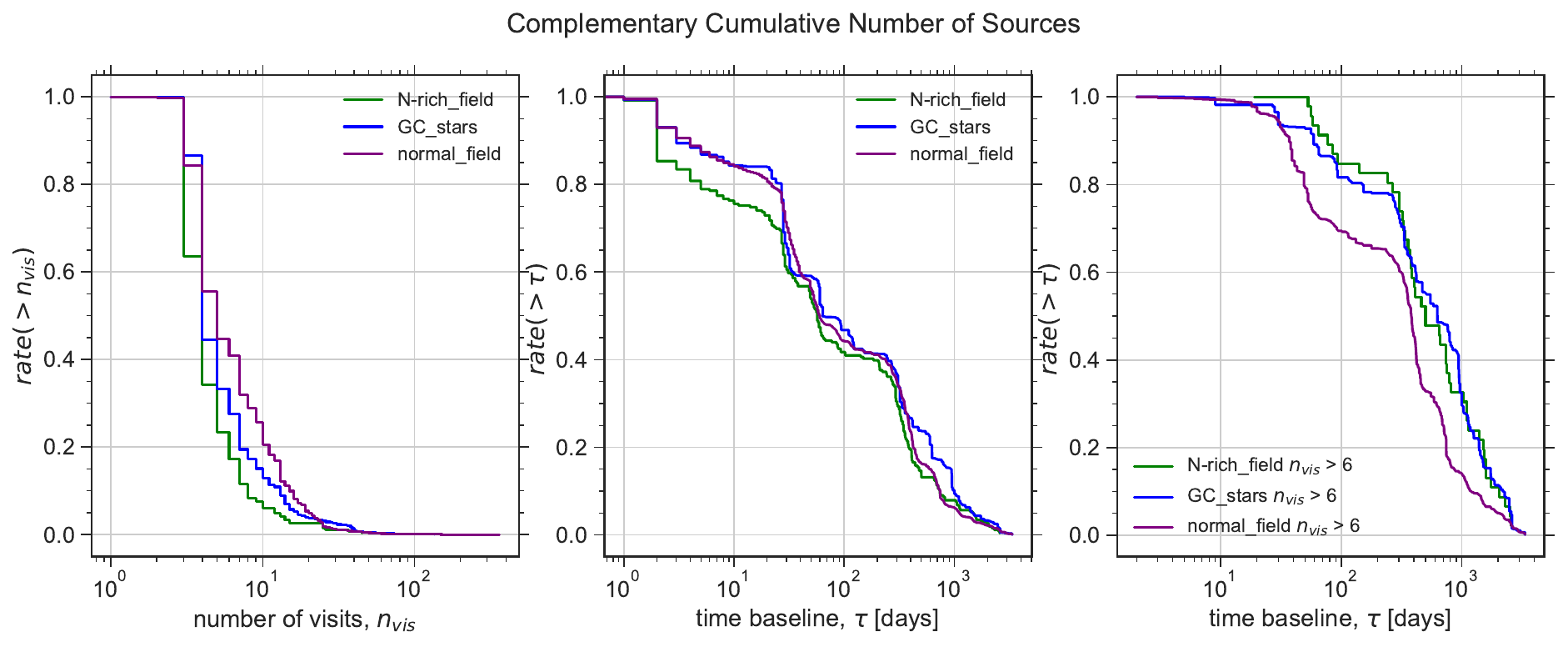}
    \caption{
    Complementary cumulative distributions of APOGEE visit statistics for the three samples: N-rich field stars (green), GC stars (blue), and normal field stars (purple). 
    (a)Left panel: fraction of stars with visit numbers larger than a given threshold. 
    (b)Middle panel: fraction of stars with temporal baselines larger than a given threshold.
    (c)Right panel:Same as panel (b), but strictly restricted to the sub-sample of stars with $\ge 6$ visits.}
    \label{fig:visit_distribution}
\end{figure}

Figure~\ref{fig:visit_distribution} presents the distributions of APOGEE visit numbers and temporal baselines for the three stellar samples analyzed in this work.
Only a minority of APOGEE targets have large numbers of visits ($N_{\rm visit}>10$), which limits the completeness of RV-based binary identification for sparsely sampled systems. In addition, most APOGEE targets have temporal baselines shorter than $\sim1000$ days, implying that the detected binaries are primarily short- and intermediate-period systems with separations of at most a few AU \citep{MoeKratter2018,Badenes2018}. Therefore, the binary fractions derived in this work correspond specifically to close spectroscopic binaries rather than the intrinsic binary fraction over all orbital separations.
Importantly, the three samples exhibit similar visit-number and baseline distributions, indicating that the APOGEE observing strategy does not introduce significant differential biases among them.

\setcounter{figure}{0} % 将图片计数器重置为0
\renewcommand{\thefigure}{B\arabic{figure}} % 定义图片编号显示格式为 A1, A2...
\section{Binary Fraction Correction}
\label{appendix:bf correction}

The spectroscopic binary fractions inferred from APOGEE radial velocities are affected by observational selection effects, including finite temporal baselines, heterogeneous visit numbers, and measurement uncertainties. To estimate the intrinsic close-binary fractions, we performed forward Monte Carlo simulations to quantify the detection efficiency and false-positive rate of our $F_2$-based method.

Our simulations are based on a parameter-matched subset of the APOGEE binary ``Golden Sample'' from \citet{PriceWhelan2020}. We strictly select systems that match the $\log g$ and $T_{\mathrm{eff}}$ parameter space of the giant stars probed in this work. We modeled the distributions of orbital eccentricity and mass ratio using Beta distributions, while the logarithmic orbital-period distribution was described by a Gaussian:

\[
f(e)=\mathrm{Beta}(e|\alpha=1.21,\beta=3.01),
\]

\[
f(q)=\mathrm{Beta}(q|\alpha=0.97,\beta=1.45),
\]

\[
\log_{10}P({\rm day})\sim \mathcal{N}(\mu=2.30,\sigma=0.57).
\]

\begin{figure}[h!]
    \centering
    \includegraphics[width=\linewidth]{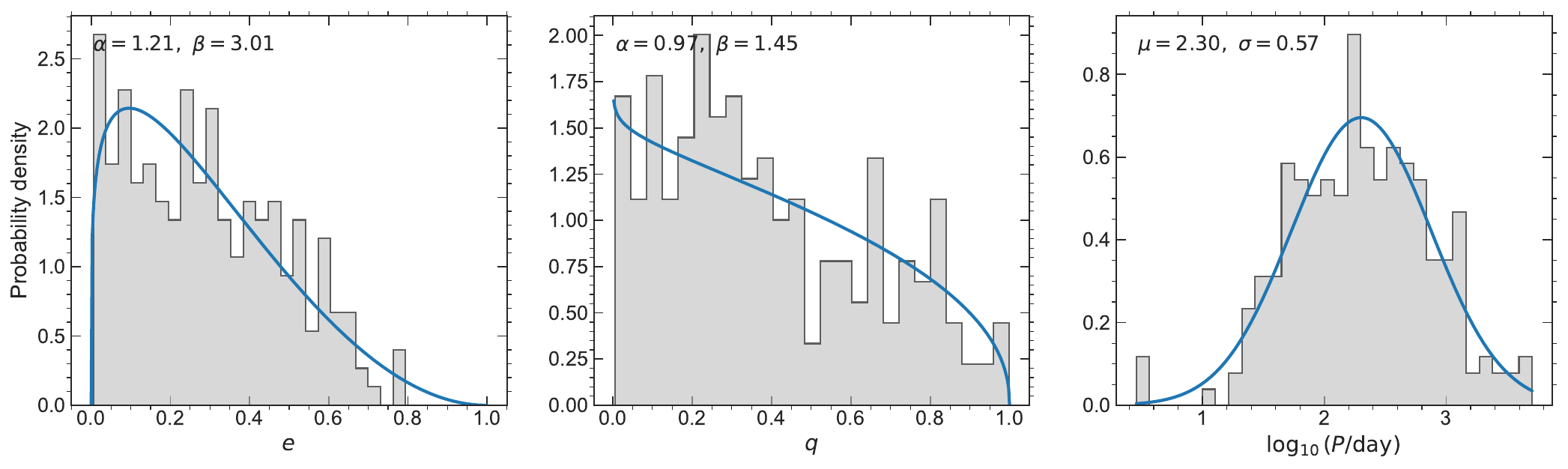}
    \caption{
    Prior distributions of binary orbital parameters derived from the parameter-matched \citet{PriceWhelan2020} ``Golden Sample''. In all panels, the grey shaded histograms represent the empirical distributions of the Golden Sample sub-population, which has been strictly filtered to match the $\log g$ and $T_{\rm eff}$ parameter space of the giant stars probed in this work. The solid blue lines denote the corresponding analytical best-fit profiles.
    \textbf{Left}: eccentricity distribution. 
    \textbf{Middle}: mass-ratio distribution. 
    \textbf{Right}: orbital-period distribution.
    }
    \label{fig:orbital_dist}
\end{figure}

The corresponding distributions are shown in Figure~\ref{fig:orbital_dist}. Remaining orbital parameters were drawn from standard isotropic distributions. We adopted a primary mass of $0.85\,M_\odot$, representative of APOGEE red giant stars.

For each independent Monte Carlo realization of the mock binary population, we generated $10^4$ mock systems, assigning half as binaries and half as single stars. Each synthetic system was sampled using the actual APOGEE visit times of randomly selected stars from the corresponding observational sample, thereby reproducing the real cadence and temporal-baseline distributions.

Mock RV curves were generated using Keplerian orbits:

\[
v_r(t)=K[\cos(\omega+\nu(t))+e\cos\omega],
\]

where the velocity semi-amplitude $K$ is

\[
K=
\left(
\frac{2\pi}{P}
\right)
\frac{a\sin i}{\sqrt{1-e^2}}
\frac{m_2}{m_1+m_2},
\]

with

\[
a=
\left[
\frac{G(m_1+m_2)P^2}{4\pi^2}
\right]^{1/3}.
\]

Applying the same $F_2>3$ criterion to the mock samples, we computed the detection efficiency $\epsilon$ and false-positive rate $f_{\rm FP}$. The intrinsic close-binary fraction was then estimated using

\[
f_{\rm obs}=f_{\rm true}\epsilon+(1-f_{\rm true})f_{\rm FP}.
\]

We find similar detection efficiencies ($\epsilon\simeq0.9$) and negligible false-positive rates ($f_{\rm FP}\sim10^{-3}$) for all three samples, indicating that differences in APOGEE cadence and temporal baseline do not significantly bias our comparison of binary fractions.

\bibliography{sample7}{}
\bibliographystyle{aasjournalv7}



\end{document}